\documentclass[prl,aps,a4paper,twocolumn,superscriptaddress,nofootinbib]{revtex4-1}

\usepackage[utf8x]{inputenc}
\usepackage{amssymb}
\usepackage{amsmath}
\usepackage{graphicx}
\usepackage{bbm}
\usepackage{psfrag}
\usepackage{latexsym}
\usepackage{hyperref}

\usepackage{xcolor}
\hypersetup{
    colorlinks=true,
    citecolor=blue,
    linkcolor=blue,
    filecolor=magenta,
    urlcolor=blue}

\newcommand{\be}{\begin{equation}}
\newcommand{\ee}{\end{equation}}
\newcommand{\bea}{\begin{eqnarray}}
\newcommand{\eea}{\end{eqnarray}}
\newcommand{\ba}{\begin{eqnarray}}
\newcommand{\ea}{\end{eqnarray}}

\newcommand{\beq}{\begin{equation}}
\newcommand{\eeq}{\end{equation}}
\newcommand{\beqa}{\begin{eqnarray}}
\newcommand{\eeqa}{\end{eqnarray}}
\newcommand{\beqar}{\begin{eqnarray*}}
\newcommand{\eeqar}{\end{eqnarray*}}

\newcommand{\eq}{\begin{equation}}
\newcommand{\eqx}{\end{equation}}
\newcommand{\eqn}{\begin{eqnarray}}
\newcommand{\eqnx}{\end{eqnarray}}

\newcommand{\eg}{{\it e.g.,}\ }
\newcommand{\ie}{{\it i.e.,}\ }

\usepackage[normalem]{ulem}

\usepackage{geometry}
\newgeometry{includefoot,left=2cm,right=2cm,bottom=1cm,top=2cm}

\begin{document}

\author{Constantin Bachas}\email{costas.bachas@phys.ens.fr}

\affiliation{\it Laboratoire de Physique  de l'\'Ecole Normale Sup\'eri{e}ure, \\
CNRS, PSL  Research University and Sorbonne Universit\'es,
24 rue Lhomond, 75005 Paris, France}

\author{Shira Chapman}\email{s.chapman@uva.nl}

\affiliation{Institute for Theoretical Physics, University of Amsterdam, \\
Science Park 904, Postbus 94485, 1090 GL Amsterdam, The Netherlands}

\author{Dongsheng Ge}\email{ge@lpt.ens.fr}

\affiliation{\it Laboratoire de Physique  de l'\'Ecole Normale Sup\'eri{e}ure, \\
CNRS, PSL  Research University  and Sorbonne Universit\'es,
24 rue Lhomond, 75005 Paris, France}

\affiliation{Kavli Institute for Theoretical Physics, University of California, Santa Barbara, CA 93106, USA}

\author{Giuseppe Policastro}\email{policast@lpt.ens.fr\\ \\}

\affiliation{\it Laboratoire de Physique  de l'\'Ecole Normale Sup\'eri{e}ure, \\
CNRS, PSL  Research University  and Sorbonne Universit\'es,
24 rue Lhomond, 75005 Paris, France}

\title{Energy Reflection and Transmission at 2D Holographic Interfaces}

\begin{abstract}
Scattering from  conformal interfaces in two dimensions is universal in that the flux of reflected and transmitted energy does not depend on the details of the initial state.
In this letter, we present the first gravitational calculation of energy reflection and transmission coefficients for interfaces with thin-brane holographic duals. 
Our result for the reflection coefficient depends monotonically on the tension of the dual string anchored at the interface, and
obeys the lower bound recently derived from the ANEC in conformal field theory.
The B(oundary)CFT limit is recovered for infinite ratio of the central charges.
\end{abstract}

\maketitle

\noindent \emph{1. Introduction.--}
Conformal interfaces are ubiquitous both in condensed-matter systems and in studies of the holographic duality. Such interfaces describe the local, scale-invariant gluing of two conformal field theories, CFT$_L$ on the left and CFT$_R$ on the right.
Examples include junctions of quantum wires \cite{Wong:1994np}, line or surface defects in the critical 2D or 3D Ising models \cite{Oshikawa:1996dj}, or the gluing of superconformal gauge theories with different couplings and/or gauge groups.
In bottom-up AdS/CFT, interfaces are often modeled by codimension-one branes anchored at the AdS boundary.
Smooth (super)gravity solutions describing  top-down embeddings in string theory are also known.
Some early papers on the subject are \cite{Karch:2000gx,Karch:2000ct,Porrati:2001gx,
DeWolfe:2001pq,Bachas:2001vj,Bak:2003jk}. Additional references will be given as we proceed.

Folding spacetime along an interface converts the latter to a conformal boundary of the product theory CFT$_L\otimes \overline{\rm CFT}_R$, where the bar indicates space reflection.\footnote{We will actually restrict our discussion to non-chiral theories, for which $\overline{\rm CFT}_R =$CFT$_R$.}
The folded theory has two energy-momentum tensors, $T_L$ and $\bar T_R$, that are separately conserved in the bulk while only their sum, $T_{\rm tot} \equiv T_L + \bar T_R$, needs to be conserved at the boundary.
What distinguishes interfaces from boundaries (and ICFTs from BCFTs) is the existence of another, relative spin-2 current $T_{\rm rel} = c_R T_L - c_L \bar T_R$\,,\footnote{This combination of the energy-momentum tensors is a conformal primary of the folded theory.} which measures the exchange of energy between left and right. Here $c_L$ and $c_R$ are the central charges of the two CFTs.
As usual, things simplify considerably in two dimensions.
In this case, it was noted  in \cite{Quella:2006de} and further analyzed in \cite{Kimura:2015nka, Billo:2016cpy, Meineri:2019ycm} that the transfer of energy across the interface is controlled by a single transmission or reflection coefficient, ${\cal T}$ or ${\cal R}$, with ${\cal T}+ {\cal R} =1$.
The purpose of the present note is to derive a formula for these coefficients in the simplest holographic-interface model.

The model consists of two AdS$_3$ slices separated by a string of tension $\sigma$.
The AdS$_3$ slices have radii $\ell_L$ and $\ell_R$,\footnote{We will work in the semiclassical limit, so the radii must be much larger than $G$.} related to the CFT central charges by the Brown-Henneaux formula $c_{L,R} = 3\ell_{L,R}/2G$ \cite{Brown:1986nw}, where $G$ is the three-dimensional Newton's constant. With no loss of generality we take $\ell_L \geq \ell_R$, so that the `false' higher-energy AdS vacuum is on the left, while the `true' AdS vacuum is on the right.
For tensions inside the interval
\bea\label{interval}
0\leq {1\over \ell_R} - {1\over \ell_L}\, \leq \, 8\pi G \sigma \, \leq \, {1\over \ell_R} + {1\over \ell_L}
\eea
the string-worldsheet geometry is AdS$_2$ corresponding to the ground state of the ICFT \cite{Karch:2000ct,Bachas:2002nz}. At the extremal values of the interval the worldsheet flattens out, \ie the AdS$_2$ radius diverges.
The lower $\sigma$ limit in \eqref{interval} actually corresponds to the Coleman-De Lucia bound \cite{Coleman:1980aw} below which the false AdS$_3$ vacuum is unstable to nucleation of bubbles. This is also the BPS bound for supergravity domain walls \cite{Cvetic:1992bf}. The upper limit, on the other hand, corresponds to the Randall-Sundrum fine-tuned tension, beyond which the string worldsheet becomes de Sitter and gets anchored on a spacelike curve of the conformal boundary \cite{Karch:2000ct}.

This model has been used as a toy model of holographic defects, in particular for calculations of holographic entanglement entropy, see \eg \cite{Azeyanagi:2007qj}.
In this letter we provide the first calculation of its transport properties.
Our main result is the following formula for the energy-transmission coefficient defined in \cite{Quella:2006de},
\bea\label{TT}
{\cal T} = {4 \over \ell_L+\ell_R} \left[ {1\over \ell_L} + {1\over \ell_R} +
8\pi G \sigma\right]^{-1}\ .
\eea
Together with the central charges, ${\cal T}$ was shown \cite{Quella:2006de} to parametrize the most general two-point functions of energy-momentum tensors allowed by the symmetries of the problem.

As explained in \cite{Kimura:2015nka,Meineri:2019ycm}, what was actually defined in \cite{Quella:2006de} is the {\it weighted-average} transmission coefficient
\bea \hskip -5mm
{\cal T} = {c_L {\cal T}_L + c_R {\cal T}_R\over c_L+c_R}\, , \ \
{\rm where}\ \
{\cal T}_{L, R} = { (c_L+c_R) {\cal T} \over 2 c_{L,R}}
\eea
are the transmission coefficients for excitations incident on the interface from the left and right, respectively. Our formula for these directional
transmission coefficients reads
\bea\label{TLR}
{\cal T}_{L, R} = {2\over \ell_{L,R}}\left[ {1\over \ell_L} + {1\over \ell_R} +
8\pi G \sigma\right]^{-1}\ .
\eea
The calculation of \eqref{TT} and \eqref{TLR} is performed by scattering surface-gravity waves in a semiclassical geometry dual to
the ground state of the ICFT.
It relies on the usual condition of no outgoing waves at the Poincar\'e horizon, whose subtle implementation we explain below.

Before describing the calculation in detail, let us comment on some salient features of our result.
First, both ${\cal T}_{L}$ and ${\cal T}_{R}$ are monotonically-decreasing functions of the
tension $\sigma$. Their maximal and minimal values (in terms of the central charges) read
\begin{equation}
\begin{split} \, \hskip -5mm
{\cal T}_{L, R}^{\rm max} = {c_R \over c_{L,R}}\
 \ ,
 \qquad
 {\cal T}_{L, R}^{\rm min} = { c_{R,L}\over c_L+c_R}\ ,
 \end{split}
 \end{equation}
 or equivalently for the average coefficients
\begin{equation}\label{Rb}
\begin{split}
 {2c_Lc_R\over (c_L+c_R)^2}\, \leq\, {\cal T} \leq {2c_R\over c_L+c_R}\ \ \
 \Longleftrightarrow \\
 \ \ \
 { c_L^2+ c_R^2 \over (c_L+c_R)^2}\, \geq \, {\cal R } \geq {c_L- c_R\over c_L+c_R}\ .
\end{split}
\end{equation}
The above lower bound on ${\cal R}$ is the same as the one following from the achronal average-null-energy condition (AANEC) in the ICFT \cite{Meineri:2019ycm}. As stressed in that reference, this lower bound is stronger than the bound imposed by reflection positivity of the Euclidean theory \cite{Billo:2016cpy}, ${\cal R } \geq ({c_L- c_R\over c_L+c_R})^2$. This shows that reflection positivity does not necessarily imply the ANEC in ICFTs.\footnote{Contrary to what happens for quantum states built by the action of local operators on the Poincar\'e-invariant vacuum of a pure CFT \cite{Hartman:2016lgu}.}

If the inequality $c_L > c_R$ is strict, both $ {\cal T}_{L}$ and $ {\cal T} $ are less than 1. Total transmission to signals incident from both sides is therefore only possible between degenerate AdS$_3$ vacua separated by a tensionless string. This is the gravitational counterpart of a topological interface.

The opposite limit of total reflection, ${\cal R}\to 1$, can only be reached by taking $c_R/c_L \to 0$, \ie by depleting CFT$_R$ of degrees of freedom, relative to CFT$_L$.
This should be contrasted with the fact that in more general ICFTs, factorizable interfaces can impose reflecting boundary conditions on each side for any values of $c_L$, $c_R$.
In our minimal holographic model, on the other hand, the transmission of energy incident from the left can be shut down only if there are no degrees of freedom in the right side. Note however that in this limit ${\cal T}_R=1$, so that the (scarce) signals incident from the right are fully transmitted to the other side.\footnote{From the perspective of the false AdS vacuum, the string looks in this limit like the end-of-the-world brane of holographic BCFT \cite{Takayanagi:2011zk,Fujita:2011fp}.
As we will see from eq.\,\eqref{lW} below,
this requires $G\ell_L \sigma$ to diverge. What is referred to as tension in \cite{Takayanagi:2011zk,Fujita:2011fp} is a finite leftover piece of $\sigma$.}

We should here stress that the transport coefficient ${\cal T}$ (or ${\cal R}$) and the ground-state entropy (the logarithm of the $g$-factor) \cite{Affleck:1991tk} are independent properties of an interface.
This is illustrated by topological interfaces in free-field models which can have arbitrarily large entropy \cite{Fuchs:2007tx,Bachas:2007td} even though their transmission coefficient is always ${\cal T} =1$. The holographic duals of such interfaces are tensionless branes \cite{Fliss_2017,Gutperle_2019}, so tension is not necessarily tied to entropy.
Entanglement entropy, which contains the ground-state entropy as a finite correction to the leading logarithmically
divergent term, has been computed in a variety of holographic ICFT models, \eg \cite{Azeyanagi:2007qj,Chiodaroli:2010ur,Jensen:2013lxa,Erdmenger:2014xya,Gutperle:2015hcv}.
It would be interesting to calculate transport coefficients in these models to see how, if at all, they are correlated with entropy.

\begin{figure}[ht]
	\centering
	\includegraphics[scale=0.45]{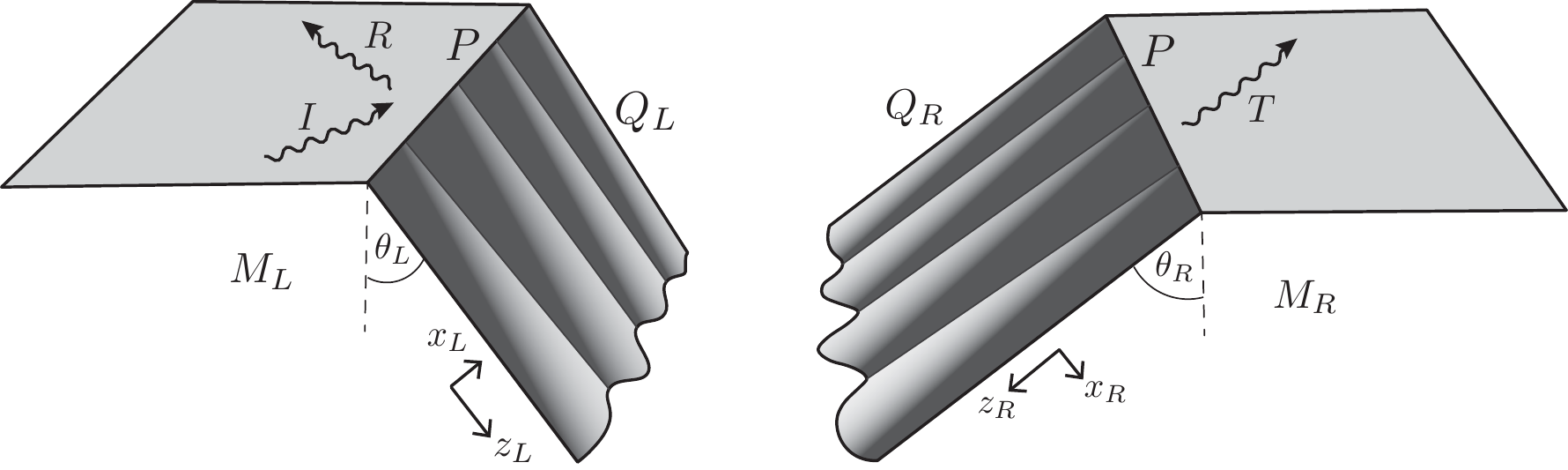}
	\caption{Illustration of the holographic-interface geometry. The two spacetimes are glued
	 together at the location of the worldsheet $Q_L \equiv Q_R$.
	 The interface $P$ is the intersection of the
	 worldsheet with the conformal boundary. The incident wave is denoted by $I$, and the reflected and transmitted waves are denoted
	 by $R$ and $T$.
	}
	\label{BraneIllustration}
\end{figure}

\vspace{15pt}
\noindent \emph{2. Holographic scattering states.--}
We describe now the main steps in the calculation of the reflection and transmission coefficients.
As mentioned above, we use a minimal holographic model for the ICFT, consisting of two manifolds $M_{L,R}$ that are locally AdS$_3$ and are joined
on the worldsheet of a tensile string. The asymptotic boundaries of these manifolds are the left, respectively right half-planes glued along the CFT interface $P$.
The latter extends in the bulk to surfaces $Q_L \subset M_L$ and $Q_R \subset M_R$ that are identified with each other and with the worldsheet of the string, see figure \ref{BraneIllustration}. The gluing of $M_L$ to $M_R$ must obey the matching conditions \cite{Israel:1966rt}
\begin{subequations}\label{israel}
\begin{align}
\gamma_{L,\alpha\beta} =& \gamma_{R,\alpha\beta} \,, \label{israel1} \\
 [K_{\alpha\beta}] - [\textrm{tr} K] \gamma_{\alpha\beta} =& 8 \pi G \sigma \, \gamma_{\alpha\beta} \,, \label{israel2}
\end{align}
\end{subequations}
where we have denoted by $\gamma_{L,R}$ and $K_{L,R}$ the induced metric and extrinsic curvature on $Q_{L,R}$, respectively, and we use $[X] \equiv X_L - X_R$ to indicate discontinuities on the two sides of the interface.

\vspace{12pt}

The ICFT
 vacuum is described in Fefferman-Graham coordinates
 by the solution \cite{Karch:2000ct,Bachas:2002nz}
\beq
\begin{aligned}
&\hskip -2mm ds^2_{L} = \frac{\ell_L^2}{y_L^2} [dy_L^2+du_L^2-dt_L^2] \ \ \ {\rm for}\, \ u_L\leq y_L \tan\theta_L ,
\\
 & \hskip -2mm ds^2_R = \frac{\ell_R^2}{y_R^2} [dy_R^2+du_R^2-dt_R^2] \, \ \ {\rm for}\,
 \ u_R\geq -y_R \tan\theta_R ,
\end{aligned}
\eeq
where $0\leq y_{L,R} < \infty$. The worldsheet $ u_L = y_L \tan\theta_L $ or $ u_R = -y_R \tan\theta_R $ subtends an angle ${\pi\over 2}+ \theta_L$,
respectively ${\pi\over 2}+ \theta_R$, to the left/right halves of the conformal boundary. The worldsheet metric is AdS$_2$ with radius $\ell_W$ obeying
\beq\label{lW}
\ell_W = {\ell_L\over \cos \theta_L} = {\ell_R\over \cos \theta_R} =
{\tan \theta_L + \tan \theta_R\over 8 \pi G \sigma} \,.
\eeq
The first two equalities follow from \eqref{israel1} and the last one from \eqref{israel2}.
It will be later convenient to employ the rotated coordinates
\begin{equation}\label{global}
\begin{split}
\begin{pmatrix} u_L\\ y_L\end{pmatrix} &= \begin{pmatrix}
\cos\theta_L & \sin\theta_L \\ -\sin\theta_L & \cos\theta_L \end{pmatrix}
\begin{pmatrix} x_L\\ z_L\end{pmatrix} \,, \ \ \ \, \\
\begin{pmatrix} u_R\\ y_R\end{pmatrix} &= \begin{pmatrix}
\cos\theta_R & -\sin\theta_R \\ \sin\theta_R & \cos\theta_R \end{pmatrix}
\begin{pmatrix} x_R\\ z_R\end{pmatrix}\ ,
\end{split}
\end{equation}
in which the unperturbed string sits at $x_L=x_R=0$, and its worldsheet can be parametrized by $t_L=t_R\equiv t$ and $z_L=z_R\equiv z$.

In principle one would like to solve the matching problem \eqref{israel} for a generic metric and a fluctuating interface on the conformal boundary. It is however sufficient for our purposes here to set all ICFT sources to zero, and only consider normalizable excitations of the fields.
These are particularly simple in pure AdS$_3$ where the most general solution of the Einstein equations in Fefferman-Graham coordinates can be written as \cite{Skenderis:1999nb} (see also \cite{Banados:1998gg,Balasubramanian:1998sn,Rooman:2000ei,Krasnov:2001cu})
\bea\label{SS}
\hspace{-16pt} ds^2 =\frac{ \ell^2 dy^2}{y^2}
+ \Bigl[ \frac{    \ell^2
g^{(0)}_{\alpha\beta} }{y^2} + g^{(2)}_{\alpha\beta} + {   y^2
\over  4 \, \ell^2 } g^{(4)}_{\alpha\beta} \Bigr] dw^\alpha dw^\beta
\eea
with $g^{(4)}= g^{(2)}(g^{(0)})^{-1}g^{(2)}$ and, for  flat boundary metric,
$g^{(2)}_{\alpha\beta} = 4 G \ell \langle T_{\alpha\beta} \rangle$. Here
$\langle T_{\alpha\beta} \rangle$ is the vev of the
canonically-normalized, traceless conserved energy-momentum tensor in some state of the dual CFT.
Linearizing in the perturbation allows us to drop $g^{(4)}$, so that the correction to the standard AdS$_3$ Poincar\'e metric has arbitrary left- and right-moving waves, $g^{(2)}_{++}(w^+)$ and $g^{(2)}_{--}(w^-)$.

In order to reproduce the setup of ref.\,\cite{Meineri:2019ycm} we consider a configuration with an incoming wave from the left,  giving rise to a reflected wave on the left  and a transmitted wave on the right.
Explicitly, identifying the $w^\pm$ of \eqref{SS} with $u\pm t$, and using monochromatic waves,\footnote{Since we are working at the linearized level, the plane wave solutions can be superposed to wave packets.} we have
\beq\label{metricp}
\begin{aligned}
  {  \bigl[ ds^2\bigr]^{(2)}_L}
   =\, & 4 G \ell_L \epsilon \Bigl[ e^{i \omega(t_L - u_L)} \, d(t_L - u_L)^2 +
\\ & \hskip 12mm {\cal R}_L\, e^{i \omega(t_L + u_L)} \, d(t_L + u_L)^2 \Bigr] + c.c.\ , \\
 {   \bigl[ ds^2\bigr]^{(2)}_R} =\,& 4 G \ell_R \epsilon \, {\cal T}_L e^{i \omega(t_R - u_R)} \, d(t_R - u_R)^2 + c.c. \, ,
\end{aligned}
\eeq
where ${\cal R}_L$ and $ {\cal T}_L$ are the (a priori complex) relative amplitudes of the reflected and  transmitted waves, and the subscript $L$ indicates that the incident wave came from the left. Anticipating the final result, we give the same names to these amplitudes as to the (real) reflection and transmission coefficients. In what follows, we will linearize our equations in the incoming flux $|\langle T_{--}\rangle|= \epsilon$. 

Gluing $M_L$ with $M_R$ requires matching coordinates on  the   worldsheet.
We allow for this by writing  $z_{L,R} = z+  {\tilde \epsilon}\, \zeta_{L,R}(z, t)$ 
 and $ t_{L,R} = t + \tilde \epsilon\, \lambda_{L,R}(z, t)$, where $z,t$ are the Poincar\'e coordinates of the AdS$_2$ worldsheet and we defined for convenience $\tilde\epsilon = {4G \over \ell_W}\, \epsilon$.
 Since we are keeping only linear order in $\epsilon$, we can set $t_L=t_R=t$ and $z_L=z_R=z$ in the perturbation \eqref{metricp}.
The above changes of coordinates enter only through the expansion of the leading worldsheet metric and extrinsic curvatures in \eqref{israel}.
We also let $x_{L,R} = \tilde \epsilon\, \delta_{L,R}(z, t)$ be the fluctuating position of the string in the transverse dimension.\footnote{We use units where the metric is dimensionless, 
 $t,x,z$\,  have dimensions of length, 
$\epsilon$ and $\tilde\epsilon$  have dimensions  of mass squared, and hence 
the functions  $\zeta_{L,R}, \lambda_{L,R},\delta_{L,R}$ have dimensions of length cubed.}

Thanks to time-translation invariance we are allowed to work at fixed frequency,
\bea
\delta_{L,R}(z, t) = e^{i\omega t} \delta_{L,R}(z) + c.c.
\eea
and similarly for $\zeta_{L,R}$ and $\lambda_{L,R}$.

We have then six equations for the six functions $\delta_{L,R}$, $\zeta_{L,R}$ and $\lambda_{L,R}$. But common reparametrizations of the two charts are pure gauge, so only the transition functions
\bea
\zeta \equiv \zeta_L - \zeta_R \quad {\rm and} \quad
\lambda \equiv \lambda_L-\lambda_R
\eea
enter in the equations \eqref{israel}.
The problem may now look overconstrained, but two of the matching conditions \eqref{israel2} are not actually independent equations. The reason is that all foliations of AdS$_3$ obey the momentum constraints
\bea\label{mom}
D^\alpha K_{\alpha \beta} - D_\beta K = 0\ ,
\eea
where $D_\alpha$ is the covariant derivative with respect to the induced metric.
Thus, once one of the equations \eqref{israel2} has been solved, the other two are
automatically satisfied up to constants.\footnote{Since the time dependence is fixed, \eqref{mom} 
implies  that the $z$ derivatives of two  matching conditions are identically zero. Note that in $D$ spacetime dimensions the same counting gives $(D-1)^2$ matching conditions for $D+1$ arbitrary functions, so that for $D>3$ two generic spacetimes cannot be matched.}

The brane fluctuations are induced by the gravity waves \eqref{metricp}.
The equations are more compact in terms of the combinations
\bea\hskip -4mm
D \equiv \delta_L - \delta_R\, , \ \ \
\Delta \equiv \tan \theta_L \delta_L +\tan \theta_R \delta_R - \zeta\, .
\eea
The four independent matching conditions read
\beq\label{4eqs}
\begin{aligned}
& \hspace{-10pt} \Delta { \, +\, i\omega z \lambda} \, =\, z^3 \Bigl[ {\cos \theta_L \over 2} ({\bf I}+
{\bf R}) - {\cos \theta_R \over 2} {\bf T} \Bigr] \,, \\
& \hspace{-10pt}  i\omega z \zeta { \, -\,
z\partial_z \lambda }\ \, =\, z^3 \Bigl[ {\sin \theta_R} \cos \theta_R {\bf T}+{\sin \theta_L } \cos \theta_L ({\bf I} - {\bf R}) \Bigr] \,, \\
& \hspace{-10pt} z\partial_z \zeta + \Delta \, =\, z^3 \Bigl[ {\sin^2 \theta_R \cos \theta_R \over 2} {\bf T} - {\sin^2 \theta_L \cos \theta_L \over 2}
({\bf I}+ {\bf R}) \Bigr] \,, \\
& \hspace{-10pt} z\partial_z D = z^3 \Bigl[ {1\over i \omega z} ({\bf I} - {\bf R}-{\bf T}) - {
\sin \theta_L\cos^2 \theta_L\over 2}({\bf I}+ {\bf R}) \\
&\hskip 3cm - {
\sin \theta_R\cos^2 \theta_R\over 2}{\bf T} \Bigr] \ ,
\end{aligned}
\eeq
where
\begin{equation}\label{IRTdef22} 
\begin{split}
&{\bf I}  \equiv e^{- i \omega \sin \theta_L z}, \
{\bf R} = {\cal R}_L \, e^{ i \omega \sin \theta_L z}, \
\\
&
~~~~~~~~~~~~~{\bf T} \equiv {\cal T}_L e^{ i \omega \sin \theta_R z}
\end{split}
\end{equation}
are the exponentials imprinted on the worldsheet by the graviton waves \eqref{metricp}.
The first three equations are the matching conditions \eqref{israel1} while the fourth is the $(tz)$ component of \eqref{israel2}, where we have used the second equation to simplify it.
The three (almost) redundant matching conditions can be actually combined into an algebraic equation for $D$, so the integration constant in the last equation of \eqref{4eqs} is fixed as in eq.\,\eqref{solutions}, see below.

Consider first the homogeneous equations obtained by setting the right-hand sides in \eqref{4eqs} to zero.
The general solution reads 
\bea\label{genhom}
& -i\omega \lambda(z) = { \Delta(z)\over z} = {a_+} e^{i \omega z}+ {a_-} e^{-i \omega z}\,, \nonumber \\
& -i\omega \zeta(z) = {a_+} e^{i \omega z}-
{a_-} e^{-i \omega z}\,, \quad D=0\ .
\eea
The $z= 0$ limit of these functions corresponds to sources in the dual ICFT.
For instance $\delta_L(0)=\delta_R(0)$ is a source for the interface displacement operator.\footnote{Similarly, $\lambda(0)$ is the source for the dual operator that generates a relative reparametrization of the interface \cite{DefectsAndFun2}.}
Linearizing in this source gives an $O(z^{-3})$ correction to the induced metric. This is consistent with the fact that the scaling dimension of the displacement operator is $\mathfrak{D}=2$ \cite{Billo:2016cpy}.
In the absence of gravity waves, setting the sources to zero implies $a_+=a_- =0$.
This shows that there are no normalizable states supported entirely by the interface.

Let's go back now to the inhomogeneous equations \eqref{4eqs}. Since these are linear equations,
the general solution is given by \eqref{genhom} plus some special solution. The result after straightforward manipulations is
\begin{widetext}
\begin{equation} \label{solutions}
\begin{split}
\frac{\Delta(z)}{z}=& \frac{1}{\omega^{2} \cos \theta_L}\left({\bf I}+{\bf R}\right)-\frac{1}{\omega^{2} \cos \theta_R} {\bf T}+a_+ e^{i \omega z}+a_- e^{-i \omega z} \,,
\\
\zeta(z)=&- \frac{\cos \theta_L z}{\omega^{2}}\left({\bf I}+ {\bf R}\right)
-\frac{i}{\omega^{3}}\left({\bf I}-{\bf R}\right)\left(\tan \theta_L+\frac{\sin \theta_L \cos \theta_L}{2} \omega^{2} z^{2}\right)
\\
&-\frac{i}{\omega^{3}} {\bf T}\left(\tan \theta_R+i \cos \theta_R \omega z+\frac{\sin \theta_R \cos \theta_R}{2} \omega^{2} z^{2}\right)+\frac{i}{\omega}\left(a_+ e^{i \omega z}-a_- e^{-i \omega z}\right) \,,
\\
\lambda(z)= &\frac{i}{\cos \theta_L \omega^{3}}\left({\bf I}+{\bf R}\right)\left(1-\frac{\cos^2 \theta_L}{2} \omega^{2} z^{2}\right)
-\frac{i}{\cos \theta_R \omega^{3}} {\bf T}\left(1-\frac{\cos^2 \theta_R}{2} \omega^{2} z^{2}\right)
+\frac{i}{\omega}\left(a_+ e^{i \omega z}+a_- e^{-i \omega z}\right) \,,
\\
D(z) =& -\frac{i}{\omega^{3}}\left({\bf I}-{\bf R}\right)\left(1+\frac{\cos^2 \theta_L}{2} \omega^{2} z^{2}\right)
+\frac{\sin \theta_L z}{\omega^{2}}\left({\bf I}+{\bf R}\right)
+\frac{i}{\omega^{3}} {\bf T}\left(1-i \sin \theta_R \omega z+\frac{\cos^2 \theta_R}{2} \omega^{2} z^{2}\right) \,.
\end{split}
\end{equation}
\end{widetext}
 Requiring that the sources vanish now gives
\bea\label{R+T}
D(0)=0 \, \Longrightarrow\,
{\cal R}_L+{\cal T}_L=1\ ,
\eea
 and further from $\zeta(0) = \lambda(0)=0$ we obtain:
\bea \hskip -9mm \label{c1}
a_+ = {1 \over 2 \omega^2} \Bigl[
{\cal T}_L \Bigl(\frac{1+\sin \theta_R}{\cos \theta_R} + \tan \theta_L \Bigr)
-\frac{(1+{\cal R}_L)}{ \cos \theta_L}
 \Bigr]\,,
\eea
\bea\hskip -9mm
a_-= {1 \over 2 \omega^2} \Bigl[
{\cal T}_L\Bigl(\frac{1-\sin \theta_R}{ \cos \theta_R} - \tan \theta_L\Bigr)
-\frac{(1+{\cal R}_L)}{ \cos \theta_L}
\Bigr]\,.
\eea
The reader can verify that with these choices all four functions are $O(z^3)$ near the conformal boundary, and make $O(1)$ contributions to the worldsheet metric which can be interpreted as ICFT vevs.
This agrees again with the fact that the scaling dimension of the displacement operator is two \cite{Billo:2016cpy}.

Inserting the solution for $\delta_{L,R}$ in the expression for the induced metric shows that the latter is locally AdS$_2$ (constant intrinsic Ricci curvature). Thus, as is the case for homogeneous AdS$_3$/CFT$_2$, here too the  dynamics happens at the conformal boundary in spite of the presence of the string/interface.


Up to this point, we have obtained a solution for the equations of motion of our model, that is
valid for any value of ${\cal T}_L$. To proceed further, we have to make an assumption about the behaviour of the solution at the Poincar\'e horizon, as mentioned in the introduction. It is well-known that in the Lorentzian AdS/CFT correspondence the boundary conditions
at the conformal boundary do not determine the solution uniquely, because there are normalizable modes that vanish at the boundary and are regular in the interior \cite{Balasubramanian:1998sn}; this is the dual of the property that there are different Minkowskian QFT propagators, depending on the choice of the initial state (retarded, advanced, Feynman etc.).\footnote{One could try to circumvent the problem by going to Euclidean signature, however in AdS$_3$ there are subtleties because one finds infrared divergences at $z\to\infty$ that must be regulated (in \cite{Skenderis:1999nb} an IR cutoff was used) and this would introduce some ambiguities.}

The prescription of \cite{Son:2002sd,Herzog:2002pc} (generalized by \cite{Skenderis:2008dg}), frequently used in the literature, requires the absence of modes coming out of the horizon for the computation of a retarded correlator. In our case it is not immediately obvious how to apply this prescription, since the problem is not formulated as the computation of a causal response.\footnote{Perhaps this can be done using an alternative definition of ${\cal T}$ in terms of a 3-point function \cite{Meineri:2019ycm}.}
One difficulty is that wave packets formed from \eqref{metricp} are localized in $u_{L,R}$ but not in the radial AdS coordinates $y_{L,R}$. Such wavepackets imprint superluminal waves on  the functions $\delta_{L,R}$, $\zeta$ and $\lambda$ of the form
$e^{i\omega t}\times ({\bf I}, {\bf R}\ {\rm or}\ {\bf T})$, see eq.~\eqref{IRTdef22}.
But as illustrated by seawaves hitting an oblique seashore, these superluminal waves carry no energy. To see why, one must look at gauge-invariant quantities left unchanged by common reparametrizations of the two charts, $\delta \zeta_L=\delta \zeta_R$ and $\delta \lambda_L=\delta \lambda_R$. One such quantity,
at the linearized order considered here, is the traceless part of the extrinsic curvature which   is continuous across the worldsheet by Israel's matching condition \eqref{israel2}.\footnote{It is also covariant under Weyl transformations of the bulk geometry \cite{Carter:2000wv}.} A simple calculation gives
\bea
\hat K_{\pm\pm} = \frac{ a_{\pm}\, \omega ^2   \epsilon}{2 \pi  \sigma \ell_W} e^{i \omega x^\pm }+\mathcal{O}(\epsilon^2) \ ,
\eea
where $x^\pm = t\pm z$ and $\hat K_{\alpha\beta}$ denotes the traceless part of $K_{\alpha\beta}$.
Note that the superluminal waves disappeared from the above expression, and
that  the `no outgoing wave' condition reduces to $a_+ =0$. Note in addition
 that the (discontinuous) trace parts, $K_{L,R}=     \pm {2\over \ell_W}
\tan(\theta_{L,R})+ \mathcal{O}(\epsilon^2)$, are not perturbed at linear order.

With the help of equations \eqref{R+T} and
\eqref{c1}, the no-outgoing-wave condition  implies
\be
{\cal T}_L = \frac{2 \cos \theta_R}{\cos \theta_R(1+\sin\theta_L)+\cos \theta_L(1+\sin\theta_R)} \,.
\ee
Trading the angles for $\ell_{L,R}$ and $\sigma$ gives our result \eqref{TLR}. It is non-trivial that ${\cal R}_L$ and ${\cal T}_L$, which started out as complex amplitudes in the gravitational-scattering problem, ended up as real, positive reflection and transmission coefficients as required for a proper ICFT interpretation.
This together with the fact that our result obeys the non-trivial ANEC bound \eqref{Rb} is a strong {\it a posteriori} argument for the correctness of the above assumption.\footnote{For instance, one can check that the condition $a_-=0$ would lead to unphysical values for ${\cal T}$.}



\vspace{15 pt}

\noindent \emph{3. Summary and Outlook.--}
In this letter we evaluated the reflection and transmission from thin-brane holographic interfaces in AdS$_3$. We found that the result \eqref{Rb} for the reflection coefficient is consistent with the lower ANEC bound, while its maximum approaches ${\cal R}=1$ only in the limit of infinite ratio of the central charges. This imperfect reflection might be a generic feature of holographic interfaces.

It would be interesting to study applications of our work in condensed matter systems, as well as explore other holographic models, higher dimensions and quantum-gravitational corrections.
Of special interest are the $1/2$-BPS holographic interfaces of $N=4$ super Yang-Mills \cite{DHoker:2007zhm,DHoker:2007hhe} and the associated top-down embedding of massive gravity \cite{Bachas:2018zmb}. 
Another important  issue that will be discussed in a future publication \cite{DefectsAndFun2}
 is universality, in particular why ${\cal R}_{L,R}$ and ${\cal T}_{L,R}$ are independent of the nature of the incident wave as has been shown in the dual  CFT$_2$  \cite{Meineri:2019ycm}.

It is also interesting to explore the relation of our work to the recent discussions of the Page curve
that describe the entanglement entropy between an evaporating black hole and its Hawking radiation through the appearance of islands behind the horizon. This has been evaluated in a class of toy models where the black hole is coupled to a heat bath via transparent boundary conditions \cite{Penington:2019npb,Almheiri:2019psf}. Holographic realizations corresponding to this scenario were put forward for example in \cite{Almheiri:2019hni,Chen:2019uhq,Rozali:2019day,Geng:2020qvw,Chen:2020uac} in terms of doubly holographic BCFT/ICFT models.
Our results on reflection and transmission could come to use when coupling the black hole to the bath -- we hope to return to this question in the future.

In this context it has been also pointed out that the transmission of energy across an interface differs from the transmission of information. It would be interesting to compare our results to various information theoretic measures and their dynamics in the presence of defects, see \eg \cite{Azeyanagi:2007qj,Sakai:2008tt,Jensen:2013lxa,Erdmenger:2014xya,Gutperle:2016gfe,Czech:2016nxc,Chapman:2018bqj}.

%
%
%
%
%
%
%
%

\begin{acknowledgments}
\section*{Acknowledgments}
We would like to thank Denis Bernard, Damian Galante, Oleksandr Gamayun, Christopher Herzog, Diego Hofman, Donald Marolf, Marco Meineri, Yaron Oz, Vassilis Papadopoulos,
Joao Penedones  and Massimo Porrati for valuable comments and discussions.
DG is grateful for the graduate fellowship program at KITP-UCSB, where part of this work was carried out. This research was supported in part by the Heising-Simons Foundation, the Simons Foundation, and National Science Foundation Grant No. NSF PHY-1748958. SC aknowledges the support of
 the ERC consolidator grant QUANTIVIOL awarded to Ben Freivogel.

\end{acknowledgments}

\vspace{10pt}

\bibliography{DefectHolo}{}


%
%


\end{document}